\documentclass[final,5p,times,twocolumn]{elsarticle}

\usepackage{amssymb}
\usepackage{amsmath}
\usepackage{xcolor}
\usepackage{soul}

\usepackage{hyperref}

\makeatletter
\long\def\MaketitleBox{%
  \resetTitleCounters
  \def\baselinestretch{1}%
  \begin{\elsarticletitlealign}%
   \def\baselinestretch{1}%
    \Large\@title\par\vskip18pt
  \ifdoubleblind
    \vspace*{2pc}%
  \else
    \normalsize\elsauthors\par\vskip10pt
    \footnotesize\itshape\elsaddress\par\vskip36pt
  \fi
    \ifvoid\absbox\else\unvbox\absbox\par\vskip10pt\fi
    \ifvoid\keybox\else\unvbox\keybox\par\vskip10pt\fi
    \end{\elsarticletitlealign}%
}
\makeatother

\journal{Science Bulletin}

\begin{document}

\begin{frontmatter}



\title{High-Pressure Crystal Structure Database}

\cortext[cor1]{Corresponding author}

\author[1,2]{Zhenyu Wang}

\author[1,2]{Qingchang Wang}

\author[1,2]{Junwen Duan}

\author[1,2]{Heng Ge}

\author[1,2]{Xiaoshan Luo}

\author[1,2]{Pengyue Gao}

\author[1,2]{Wei Zhang\corref{cor1}}
\ead{zhangw_bxx@jlu.edu.cn}

\author[1,2]{Jian Lv\corref{cor1}}
\ead{lvjian@jlu.edu.cn}

\author[1,2]{Yanchao Wang\corref{cor1}}
\ead{wyc@calypso.cn}

\author[4,5,1,2,3]{Yanming Ma\corref{cor1}}
\ead{mym@jlu.edu.cn}

\affiliation[1]{organization={Key Laboratory of Material Simulation Methods and Software of Ministry of Education, College of Physics, Jilin University},
            city={Changchun},
            postcode={130012}, 
            country={China}}

\affiliation[2]{organization={State Key Laboratory of High Pressure and Superhard Materials, College of Physics, Jilin University},
            city={Changchun},
            postcode={130012}, 
            country={China}}
            
\affiliation[3]{organization={Changbaishan Laboratory and International Center of Future Science, Jilin University},
            city={Changchun},
            postcode={130012}, 
            country={China}}

\affiliation[4]{organization={Center for High-Pressure Science and Technology, Zhejiang University},
            city={Hangzhou},
            postcode={310027}, 
            country={China}}
            
\affiliation[5]{organization={School of Physics and Institute of Fundamental and Transdisciplinary Research, Zhejiang University},
            city={Hangzhou},
            postcode={310027}, 
            country={China}}

\end{frontmatter}

Pressure is a fundamental thermodynamic variable that reshapes phase stability and material properties. As the $PV$ term grows in the enthalpy ($H = E + PV$), compression reorganizes chemical bonds, modifies atomic coordination, alters electronic states, and drives structural phase transitions \cite{HighPressureReview2017NRM}. Building on this principle, advances in static ultrahigh-pressure experiments, \textit{in situ} characterizations, and computational crystal structure prediction (CSP) have turned high-pressure research into a productive route to new structures and emergent properties. A striking example is the discovery of hydrogen-rich high-temperature superconductors: CSP first identified key clathrate-hydride candidates \textit{in silico} \cite{CaH6The2012PNAS,LaH10The2017PNAS,PengHydride2017PRL,LaSc2H24The2024PNAS}, guiding subsequent synthesis and pushing the superconducting transition temperature from $\sim$215~K in CaH$_6$ \cite{CaH6Exp2022PRL,CaH6Exp2022NC} and $\sim$250~K in LaH$_{10}$ \cite{LaH10Exp2019Nature,LaH10Exp2019PRL} to 271--298~K in LaSc$_2$H$_{24}$ \cite{LaSc2H24ExpArxiv} at several hundred gigapascals.

Despite the rapidly surging number of theoretically predicted and experimentally realized high-pressure phases, the infrastructure required to systematically manage, standardize, and mine this expanding configuration space has severely lagged. In ambient-pressure materials science, high-throughput first-principles calculations and comprehensive databases—such as the Materials Project \cite{MP2013APLM} and OQMD \cite{OQMD2013JOM}, which standardize structural and stability data at 0~K and 1~atm—have revolutionized the field by enabling large-scale consistency and data-driven discovery. By contrast, public datasets dedicated to high-pressure materials remain sparse and disjointed, highlighting a critical infrastructure gap in extreme-condition research.

\begin{figure*}[t]
\centering
\includegraphics[width=0.8\linewidth]{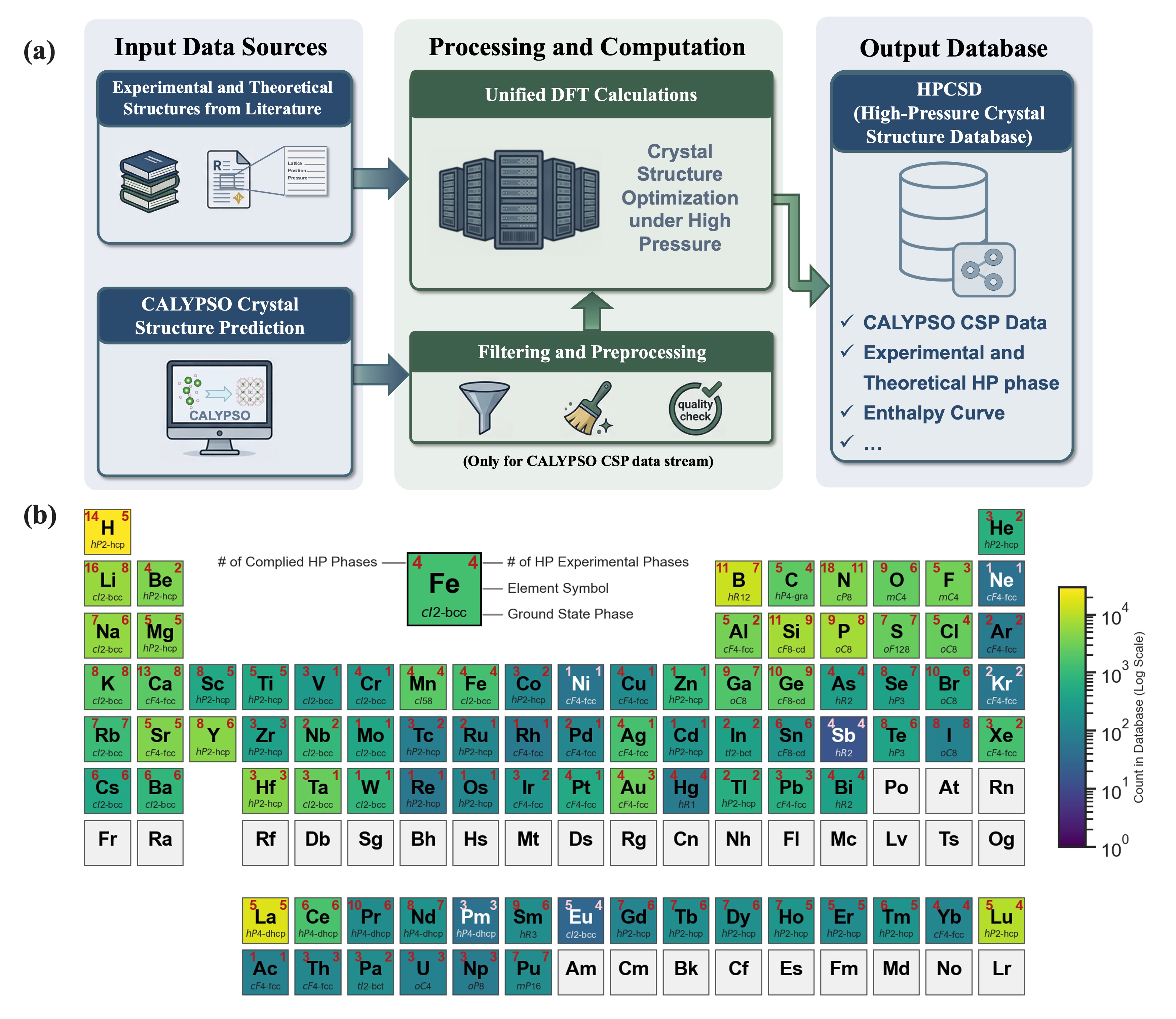}
\caption{Database construction workflow and elemental coverage of HPCSD. \textbf{(a)} Schematic overview of the database generation process. Data is sourced from two parallel streams: experimentally and theoretically reported elemental high-pressure phases from the literature, and structures generated via CALYPSO crystal structure prediction. Following rigorous screening and unified DFT optimization, the standardized entries are archived alongside their pressure-resolved structural parameters, energies, forces, and associated metadata. \textbf{(b)} Periodic table heat map illustrating the elemental distribution within HPCSD on a logarithmic scale. Each elemental cell details the ambient-pressure ground-state phase (annotated by its Pearson symbol), the number of experimentally reported elemental high-pressure phases (upper-right corner), and the total number of elemental high-pressure phases archived in HPCSD (upper-left corner). The background color illustrates the total volume of archived structures involving that element, displayed on a logarithmic scale.}
\label{fig:Fig1}
\end{figure*}

\begin{figure*}[!t]
\centering
\includegraphics[width=1.0\linewidth]{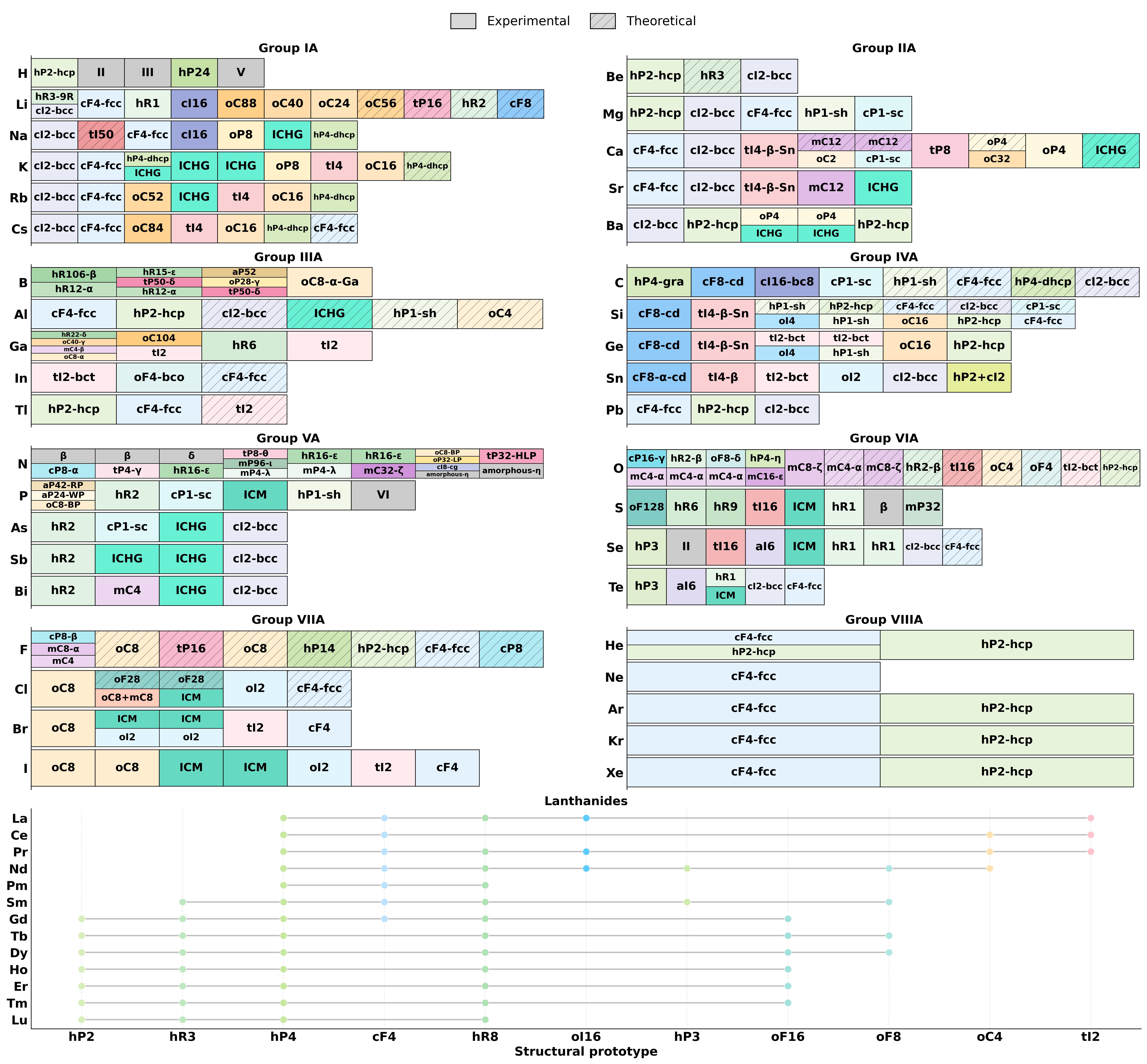}
\caption{Pressure-induced phase-transition sequences for main-group elements and lanthanides. The trajectories are summarized from experimental and theoretical data compiled in HPCSD (excluding Eu and Yb from the lanthanide series). Solid boxes denote experimentally reported phases, generally derived from room-temperature measurements, whereas hatched boxes indicate structures predicted by 0 K calculations. Pearson symbols and conventional structural labels (e.g., cF4, bcc, phase I, $\alpha$) are retained for clarity. Gray-shaded regions denote phases for which no well-defined crystal structure has been established, including disordered or amorphous states, as well as phases with unresolved or debated structural assignments. The abbreviations ICM and ICHG denote incommensurately modulated and incommensurate host–guest structures, respectively. Exact transition pressures are omitted for visual readability. A few phases predicted at ultrahigh pressures fall outside the current scope of HPCSD and are therefore absent from this figure, while certain decompression-recovered phases, though archived in the database, are also not depicted here; both categories are documented in Table S4. The vertical arrangement of boxes within an elemental sequence is schematic and is not intended to rigorously represent thermodynamic phase relations, particularly for metastable pathways or where spatial constraints required simplification.}
\label{fig:Fig2}
\end{figure*}

Consequently, crucial high-pressure structural information remains highly fragmented across individual publications and heterogeneous computational repositories. Essential data, including continuous structural evolution data, robust enthalpy-pressure relations, and rigorously defined phase boundaries, are widely scattered. This fragmentation creates a major bottleneck for data-driven materials design, as the absence of large-scale, standardized structural datasets prevents systematic trend mining and rigorous cross-study thermodynamic comparisons. While recent efforts like the HEX database \cite{HEX2024SciData} represent a valuable step forward for high-pressure elemental structures, their sparse sampling across a few discrete pressure points is prone to overlooking transition-dense regions and phases with narrow stability windows.

To bridge this gap, we introduce the HPCSD (High-Pressure Crystal Structure Database), a traceable, pressure-resolved repository that integrates experimental and theoretical high-pressure structures. As illustrated in Fig.~\ref{fig:Fig1}(a), HPCSD is constructed from two complementary data streams that play distinct but synergistic scientific roles. At present, this initial release encompasses a total of 77,346 consistently evaluated structural entries spanning 89 elements. By providing standardized, reusable, and rigorously evaluated high-pressure structural data (with detailed computational methodologies provided in the Supplementary Materials Section 1), HPCSD establishes a robust infrastructure to accelerate both experimental phase identification and theoretical exploration in extreme-condition materials science.

The first stream focuses on elemental high-pressure phases, containing both experimentally observed and theoretically predicted structures. For this branch, we totally extracted crystallographic information for 467 high pressure elemental phases, including 362 reported experiment elemental phases. To ensure rigorous comparability across all entries, the structures were re-evaluated under a unified density functional theory (DFT) framework to generate continuous enthalpy curves across their respective stability fields. By default, the majority of these enthalpy calculations were performed within a 0 to 300 GPa pressure range using a standard 10 GPa grid, with finer pressure sampling adopted near phase transitions (critical enthalpy crossings). If a structure could no longer be maintained during optimization at pressures far outside its stability field, that data was discarded, while calculations for a small subset of phases were successfully extended beyond 300 GPa. This systematic curation yielded 10,993 data entries.

The second stream, the CALYPSO branch, expands the searchable configuration space of stable and metastable phases using CSP data. This branch comprises structures from diverse chemical systems generated via community-contributed CALYPSO tasks. Because the original calculations often varied in computational settings and contained duplicate or unphysical candidates, the raw structures underwent rigorous screening and cleaning. All retained candidates were subsequently re-optimized under the same unified DFT protocol applied to the first branch, retaining only the 50 lowest-energy structures per composition. The pressures spanned by these CSP structures range from tens of GPa up to the TPa regime, reflecting the broad scope of the contributing CALYPSO studies. This computational effort currently contributes 66,353 standardized data entries.

Figure~\ref{fig:Fig1}(b) provides a periodic-table overview detailing the elemental coverage and data distribution within the current release of HPCSD. Within the map, each chemical symbol is annotated with its ambient-pressure ground-state structure, denoted by its Pearson symbol and, where applicable, its conventional structural designation (e.g., bcc for body-centered cubic). The integer in the upper-right corner of each cell indicates the number of experimental elemental high-pressure phases archived in HPCSD, whereas the upper-left corner shows the total number of elemental high-pressure phases, including both experimental and theoretical entries, for that element. Detailed high-pressure phases and related references are listed in Table S4. Additionally, the background color represents the total count of all structures in HPCSD that contain that element—spanning unary, binary, and multinary compositions—displayed on a logarithmic scale.

An analysis of the elemental phase counts reveals that pressure-induced polymorphism is ubiquitous, averaging approximately 4.1 experimental and 5.2 total (experimental and theoretical combined) elemental high-pressure phases per element across the database. Crucially, this polymorphism is not randomly distributed; instead, it exhibits pronounced family-dependent trends. Elements within the s-block, p-block, lanthanide, and actinide series demonstrate extensive phase diversity. Nitrogen exemplifies this complexity: of 17 experimentally reported high-pressure phases, 11 well-defined crystalline structures are archived in HPCSD with DFT-calculated data, while the remaining 6 disordered or amorphous states are cataloged in Table~S4. Together with 7 theoretically predicted phases such as $I\bar{4}3m$-N$_{10}$ \cite{Wang_N_2012} that have yet to be synthesized, HPCSD archives 18 elemental nitrogen structures in total. Conversely, most transition metals and noble gases, such as nickel and neon, resist structural changes under compression. This disparity demonstrates that structural diversity under compression is strongly influenced by an element's electronic adaptability. Polymorphism tends to be enhanced in systems prone to pressure-induced orbital mixing, valence delocalization, or the breakdown of directional bonds \cite{HighPressureReview2020NRC}, whereas robust, closed-shell or stable partially-filled systems generally adopt simpler dense packings.

Furthermore, the logarithmic heat map reveals a highly concentrated distribution in the total volume of archived structures per element, a feature predominantly driven by the CALYPSO CSP branch. Elements such as H, La, B, and Lu contribute a dominant share of the database, with 29,427, 16,986, 12,781, and 9,428 structures containing these elements, respectively. This mirrors the contemporary frontiers of CALYPSO high-pressure CSP research. In recent years, computational efforts have been heavily focused on these specific elements, spurred by the intense pursuit of novel high-temperature superconductors—particularly hydrogen-rich clathrate hydrides—and exotic borides. Consequently, the database naturally reflects the theoretical community's prioritized exploration of these highly promising chemical spaces.

Figure S1 provides a statistical overview of the structural, compositional, and pressure distributions within HPCSD. Most structures contain fewer than 40 atoms per primitive cell, with a distribution peak at approximately 10 atoms. This cell-size distribution closely mirrors that of the Materials Project \cite{MP2013APLM}, while being shifted toward slightly larger structures compared to the HEX database \cite{HEX2024SciData} (Fig. S1a). Compositionally, the collection is heavily dominated by binary and ternary systems, with quaternary compounds making up only a minor fraction (Fig. S1b). Furthermore, the majority of entries are concentrated between 200 and 400 GPa (Fig. S1c), directly mirroring the primary focus of the underlying CALYPSO CSP studies. However, a valuable subset of data extends into the ultrahigh-pressure regime (1–2 TPa), effectively expanding the searchable configuration space toward extreme compression.

The current release of HPCSD provides comprehensive coverage of high-pressure elemental phases alongside a vast repository of stable and metastable structures derived from CSP. While future efforts will continuously integrate experimentally reported multinary compounds and further expand the CSP library, this near-complete mapping of elemental systems already serves as a unique platform for systematic, large-scale analysis. To this end, Fig. \ref{fig:Fig2} illustrates the pressure-induced phase-transition sequences of main-group elements and lanthanides. By plotting experimental (solid bars) and theoretical (hatched bars) phases in a unified sequence, this visualization transforms fragmented literature into continuous structural trajectories. These trajectories reveal family-dependent patterns in structural evolution under compression, which can be examined more closely by analyzing individual regions of the periodic table.

Among the $s$-block elements, alkali and alkaline-earth metals exhibit rich structural diversity under compression, as depicted in Fig. \ref{fig:Fig2}. Rather than maintaining their ambient close-packed (fcc, hcp) or bcc lattices, many of these elements undergo extended intermediate-pressure sequences. These pathways feature complex phases—such as $cI16$, $oC52$, $oC88$, and intricate incommensurate host–guest (ICHG) structures—demonstrating that a low valence electron count does not guarantee structural simplicity at high pressures. Physically, this complexity arises from pressure-induced electronic redistribution, such as $s \rightarrow p$ or $s \rightarrow d$ orbital hybridization and charge transfer\cite{HighPressureReview2017NRM,HighPressureReview2020NRC}. As core electrons begin to overlap, valence electrons are frequently forced into interstitial lattice voids, fundamentally altering the elements' metallic nature and occasionally resulting in electride states\cite{MaNa2009Nature}. Beyond the $s$-block, $p$-block elements—particularly those in Groups 15 through 17—display profoundly intricate phase behaviors dictated by the competition between directional covalent bonding and packing efficiency. Elements such as N, O, I and P undergo prolonged phase-transition sequences, initially adopting element-specific intermediate states like molecular (e.g., N-$tP4$-$\gamma$, O-$mC16$-$\epsilon$, I-$oC8$-VI) or layered (e.g., P-$hR2$-II) motifs. These intermediate structures preserve clear signatures of their low-pressure chemistry. As pressure increases, the energetic cost of maintaining open, directional bonds is outweighed by the $PV$ term in the enthalpy, driving a progressive breakdown of these networks into denser polymeric, atomic, or metallic states (e.g., N-$cI8$-cg, P-$cP1$-III-sc). In stark contrast, the noble gases display highly constrained structural responses, generally maintaining simple close-packed geometries (fcc, hcp) across vast pressure ranges. Because of their highly stable, closed-shell electron configurations, noble gases lack the readily accessible orbitals required for pressure-induced hybridization, defaulting instead to pure steric packing. 

Contrasting with main-group structural diversity, the lanthanides display a highly regular high-pressure phase-transition sequence (Fig. \ref{fig:Fig2}, lower panel). Despite exhibiting diverse ground states at ambient pressure, their compression pathways rapidly converge onto a universal sequence of shared structural prototypes, aligning with recent findings \cite{LaUnified2025RPL}. Physically, this universal structural trajectory is primarily governed by a common electronic mechanism: the continuous, pressure-induced $s \rightarrow d$ charge transfer. Under compression, the broad $6s$ band rises in kinetic energy relative to the $5d$ band, steadily increasing the $d$-band occupancy\cite{LanthanidesElectronHybride1977PRL}. This continuous transfer dictates the canonical close-packed phase sequence observed across the series (e.g., hcp $\rightarrow$ Sm-type $\rightarrow$ dhcp $\rightarrow$ fcc $\rightarrow$ dfcc). At even more extreme pressures, the deeply bound, highly localized $4f$ electrons begin to delocalize and participate in metallic bonding, a phenomenon that triggers volume collapses and the emergence of lower-symmetry phases\cite{LanthanidesFelectrion2000Hyperfine}. Because the lanthanide series, with the exceptions of Eu and Yb, shares this fundamental electronic architecture, their structural evolution under pressure tends to follow a remarkably predictable and unified pathway.

Besides, the Fig.\ref{fig:Fig2} can also point out that the greatest structural complexity emerges at intermediate rather than at the highest pressures accessed experimentally in many elemental systems. It is in this crossover regime that ICHG, and incommensurate modulated (ICM) structures recur most frequently, as illustrated by Na-$tI19$, Sc-$I4/mcm(00\gamma)$, or I-$Fmmm(00\gamma)s00$.
These incommensurate phases are the natural consequences of the competition between denser packing and residual characteristics (directional bonding, band-structure, or lattice instabilities) inherited from lower-pressure phases\cite{ICReason2003Chem.AEur.J.}. Once these characteristics are suppressed, simpler structures (e.g. Na-$hP4$-dhcp, I-$cF4$-IV) often re-emerge at higher pressure. Most of ICHG and ICM structures have been summarized in Table S2 and Table S3, respectively.

Although high pressure often stabilizes unexpected structures and enriches elemental phase diversity, some structural prototypes recur in different elements across different blocks of the periodic table. For example, the commonly observed bcc structure has been found in diverse elements spanning the $s$-, $p$-, $d$-, and $f$-blocks (e.g., Li, As, Ti, and U) , while the simple hexagonal (sh) structure has been identified in several main-group elements, namely Mg, Si, P, and Ge. The hcp structure has also been extensively observed across the periodic table, appearing in systems as varied as occurring in chemically diverse elements such as H, noble gases, transition metals, lanthanides, and heavy main-group elements. Taken together, these examples indicate that several structural prototypes are not isolated cases, but recurrent motifs in elemental systems under compression. A summary of these commonly observed structural prototypes and their associated elements is provided in the Table S1.

Overall, HPCSD establishes a reusable, pressure-resolved data infrastructure for high-pressure materials research. By integrating experimentally reported phases with uniformly recalculated CSP structures, the database enables more efficient phase identification, cross-element comparison, and data-driven analysis under compression. We expect HPCSD to support both experimental exploration and theoretical discovery, and to serve as a valuable foundation for emerging data-centric approaches in the high-pressure community. Besides, its standardized structural and energetic information should facilitate the development of machine-learning interatomic potentials for high-pressure systems, where reliable training data remain limited, and provide a chemically diverse reference space for generative models aimed at structure prediction and inverse design under compression. As new structures, denser pressure grids, and richer metadata are incorporated, HPCSD will continue to expand in both coverage and utility, further strengthening its role as a bridge between first-principles calculations, data-driven modeling, and high-pressure experiments.

\section*{Conflict of interest}
The authors declare that they have no conflict of interest.

\section*{Acknowledgements}
This work was supported by the Scientific and Technological Innovation Project of Changbaishan Laboratory, Jilin Province (Grant No. CBS2025007-01), the National Natural Science Foundation of China (Grant Nos. 12374005, T2225013, 52288102 and U22A2098), and Fundamental and Interdisciplinary Disciplines Breakthrough Plan of the Ministry of Education of China (Grant No. JYB2025XDXM403). Z.W. was supported by the Postdoctoral Fellowship Program of CPSF (Grant No. GZC20252243). Part of the calculation was performed in the high-performance computing center of Jilin University. The authors sincerely thank the CALYPSO community contributors for supplying the structural data used to construct the database.

\section*{Author contributions}
Z.W. performed the data collection, DFT calculations, data analysis, and prepared the figures. Q.W., J.D., H.G., X.L., and P.G. contributed to the data collection and DFT calculations. J.L., Y.W., and Y.M. conceptualized, designed, and led the research. All authors participated in the discussions and contributed to the writing of the manuscript.

\section*{Data availability}
The database is publicly available at http://www.calypso.cn. Additional data that support the findings of this study are available from the corresponding author upon reasonable request.

\appendix
\section{Supplementary material}
\label{sec:appendix}
Supplementary data to this article can be found online at
https://doi.org/xx.xxxx/xx

\bibliographystyle{elsarticle-num-names}
\bibliography{main}

\end{document}